# Least-cost diets to teach optimization and consumer behavior, with applications to health equity, poverty measurement and international development

This version revised 18 Dec 2023, for resubmission to the *Journal of Economic Education*

**Jessica K. Wallingford**[1] and **William A. Masters**[2*]

1. PhD candidate, Friedman School of Nutrition Science and Policy, Tufts University
2. Professor, Friedman School of Nutrition Science and Policy & Dept. of Economics, Tufts Univ.
* Contact author: william.masters@tufts.edu

**Abstract:** The least-cost diet problem introduces students to optimization and linear programming, using the health consequences of food choice. We provide a graphical example, Excel workbook and Word template using actual data on item prices, food composition and nutrient requirements for a brief exercise in which students guess at and then solve for nutrient adequacy at lowest cost, before comparing modeled diets to actual consumption which has varying degrees of nutrient adequacy. The graphical example is a "three sisters" diet of corn, beans and squash, and the full multidimensional model is compared to current food consumption in Ethiopia. This updated Stigler diet shows how cost minimization relates to utility maximization, and links to ongoing research and policy debates about the affordability of healthy diets worldwide.

**Keywords:** Excel; least cost diet; linear optimization; nutrient adequacy

**JEL codes:** A2, I12, Q18

**Online materials:** The Excel workbook to implement the exercise described in this paper is available at https://sites.tufts.edu/foodecon/least-cost-diet-exercise-for-nutrient-adequacy**.**

**Acknowledgments:** This work was developed for instructional purposes at Tufts University. Jess Wallingford reports financial support from a Friedman Nutrition and Citizenship Fellowship from Tufts University, and a Social Sciences and Humanities Research Council (SSHRC) Doctoral Fellowship from the Government of Canada. We are grateful to the hundreds of students at Tufts University who completed this exercise and contributed to its development, and the dozens of students and former teaching assistants who gave detailed feedback, especially Leah Costlow, Amelia Finaret, Dierdre Schiff, and Lucy Toyama. Developing this classroom exercise brought insights that underpin the Food Prices for Nutrition project described in the paper, and we are grateful to the many project collaborators for making these ideas come to life for global food security monitoring at the FAO, the World Bank, and other agencies around the world. We also thank the *JEE* editors and three anonymous referees whose suggestions greatly improved this paper.

# Least-cost diets to teach optimization and consumer behavior, with applications to health equity, poverty measurement and international development

## Introduction and motivation

Soon after the discovery of essential nutrients in the early 20[th] century, George Stigler (1945) described the challenge of choosing a set of foods to meet those needs at lowest total cost. Stigler saw the least-cost diet as just one instance of a much larger class of multidimensional optimization problems that give insight into decision-making, and soon thereafter George Dantzig developed the simplex algorithm to find exact solutions and advance linear programming for other problems as he later described in an autobiographical essay (Dantzig 1990). The diet problem continued to be widely used in research and teaching for operations research (Garille and Gass 2001), and is used in economics research to convey the idea of a subsistence constraint on well-being for poverty measurement (Allen 2017) with many practical applications for international agriculture, food and nutrition policy (Masters et al. 2018; Herforth et al. 2020; FAO 2023, World Bank 2023).

This paper describes use of an Excel workbook to explore least-cost diet calculations and compare the results to actual food consumption choices. The workbook is available at https://sites.tufts.edu/foodecon/least-cost-diet-exercise-for-nutrient-adequacy, and is pre-populated with updated nutrient requirements for college age women and men, as well as the prices and nutrient composition of actual food items for sale in Boston Massachusetts near the start of the most recent semester in November 2023. The spreadsheet guides students through guesswork that builds intuition about optimization as in Stigler's original framing, and then reveals an exact solution to show the value of linear programming as a complement to intuition. Early versions of the diet problem often had only lower-bound constraints on a few nutrients so their solution included few foods in extreme quantities, while more recent nutrition research identifies both upper and lower bounds on a wider range of nutrients that lead to more diverse and palatable diets (Garille and Gass 2001).

For this exercise we use the most recent evidence assembled for the U.S. and Canada by the National Academies of Science, Engineering and Medicine (NASEM 2019), providing a total of 21 lower bounds and 16 upper bounds on 22 different nutrients plus energy balance. The resulting diet is a set of familiar foods that students can imagine eating, although it would fall short of diet quality criteria beyond nutrient adequacy such as those specified in the Dietary Guidelines for Americans (USDA and HHS 2020). The exercise helps students compare the stylized nutrient-adequate diet to foods that have other desirable attributes, and to diets that are actually chosen. As example benchmarks the exercise includes survey data on foods actually consumed by the poorest quintile of people in Ethiopia, and the graphical example is compared to the "three sisters" diet of Mesoamerica. Instructors and students can readily update, adapt and expand the spreadsheet to teach additional aspects of the problem, but the basic exercise can be done in a single class session that helps build students' analytical skills and familiarity with economic principles, data sources and empirical methods to understand consumer choice relative to biological requirements for health. Instructors and students interested in learning more about the topic can consult the most recent *Handbook of Agricultural Economics* review chapter on Economics of Malnutrition (Masters, Finaret and Block 2022), or the new textbook on *Food Economics: Agriculture, Nutrition and Health* (Masters and Finaret 2024).

The spreadsheet and writing exercise described here links directly to ongoing use of least-cost diets to inform dietary recommendations and aid to low-income people in high-income countries such as the Thrifty Food Plan in the U.S. Supplemental Nutrition Assistance Program (USDA 2021), or by international agencies and national governments in low-income countries such as the World Food Programme's Fill the Nutrient Gap activities (WFP 2022). More recently, least-cost diets for nutrients and food groups have been introduced as a new kind of price index to track changes in the cost of foods for health (Masters et al. 2018; Bai et al. 2020; FAO 2023; World Bank 2023), Least-cost diets are particularly useful to quantify the biological constraints on food needs in an absolute poverty line for international comparison (Allen 2017) and economic history (Moatsos 2021) as well as targeting agricultural programs and food policy (FAO, IFAD, UNICEF, WFP and WHO 2022, World Bank 2023).

The least-cost diet exercise provides an unusually accessible introduction to optimization models and consumer decision-making because all students make food choices every day. Some students are initially attracted to the exercise as a possible guide to inform their own choices: they may be familiar with the idea of nutrient requirements, and curious about what combination of foods would meet their needs in the most affordable way. Many students are concerned about social equity, asking whether low-income people are able to meet their nutritional needs. And all students can use the exercise to learn economics, distinguishing between income, prices, and preferences as causes of quantity consumed. This exercise reveals how some people lack sufficient income for nutrient adequacy, but also that the nutritional value of each item and the food combinations needed for health are credence attributes that consumers often misunderstand, and in any case consumers often pursue goals other than just their own long-term health. Topics for discussion include the origins of our diverse preferences and when absolute versus relative poverty lines might be most useful to measure deprivation and target social assistance, which can lead to very rich classroom debate (Diduch 2012).

The least-cost diet exercise uses real biological data to specify what preferences would be if people wanted only the nutrients they need for health, showing how a stylized economic model can help explain underlying similarities and drive discussion about variation in human behavior. The exercise presented here uses Excel and a Word template to make the problem accessible for students with limited previous exposure and often some anxiety about mathematics, allowing them to explore the workbook and see for themselves how the model uses familiar data to generate surprising results. As Barreto (2015) notes, the use of Excel as a pedagogical complement to standard lecturing can provide visual and tactile stimulation, and allow for repeated practice with feedback and active learning to improve outcomes in ways that are emphasized by Batt et al. (2020).

The Excel and Word templates used for this exercise provide a biologically accurate updating of the least-cost diet problem, in a form that has evolved over more than a decade to be attractive for students even if they have little or no relevant prior experience, while building and rewarding advanced skills for students who are already familiar with the topic. Every parameter of the model is visible on screen, and students can immediately see the consequences of changing each decision variable with the aid of bar charts and colored cells. The formula in each cell can also be seen but the math itself is unobtrusive. This helps students discuss the data and model structure in terms that are concrete and familiar to them, leaving matrix notation and algebra to be taught in other classes. This use of Excel to teach optimization complements its use for other numerical calculations such as population pyramids for demographic projections (Barreto 2018) and Lorenz curves for measuring inequality (Halliday 2019).

How food choice relates to least-cost diets provides a valuable introduction to consumer behavior, and a starting point for advanced work in health economics, food systems and access to healthy diets. Nutritional requirements for survival and health are a widely recognized foundation for human development, establishing a floor of basic needs that underlies individual choice and both the efficiency and equity of societal outcomes (Bowles et al. 2017). This exercise reveals how students' food choices relate to their own nutrient requirements, combining items to meet recommended daily intake of essential nutrients within energy balance. Nutrient adequacy is just one step towards overall healthy diets, which are typically specified in terms of food groups as in the Dietary Guidelines for Americans (USDA and HHS 2020) or other food-based dietary guidelines (FAO, IFAD, UNICEF, WFP and WHO 2020) used to calculate the cost and affordability of healthy diets globally (FAO 2023; World Bank 2023).

In summary, this exercise provides Excel and Word templates for a classroom exercise in which students: (1) apply multidimensional linear optimization to a familiar kind of everyday decision, (2) compare heuristic to algorithmic results, using data visualization to guide guesswork, (3) contrast least-cost modeled diets to actual food choices, and in so doing (4) gain familiarity with economic data from national accounts and household consumption surveys.

**A graphical introduction to multidimensional optimization**
This section provides a biologically accurate introduction to the diet problem in two and then three dimensions, to help build students' understanding of linear programming and constrained optimization more generally. The exercise is self-contained so instructors can open the Excel workbook directly, but sketching this section on a whiteboard or using prepared slides to show this simplified version of the full exercise can help build intuition about how least-cost diet modeling relates to actual food consumption.

***Constructing an empirically meaningful diet problem with just two or three foods***
To begin in just two dimensions, we ask what quantities of two foods (black beans and butternut squash) would be needed to meet requirements of two nutrients (vitamin A and iron). This reveals that adding a third food, for example corn flour in the form of tortillas, is needed to maintain energy balance within upper and lower bounds on each nutrient.

Complementarity between these three types of food underlies many world cuisines, with this specific example of corn, beans and squash being the "three sisters" that have long been central to the lives of Indigenous Peoples across the Americas (Marsh 2021, Ngapo et al. 2021). Additional dietary diversity would be needed to meet all nutritional needs for lifelong health, but a simple three-food diet at least cost allows students to see how optimization against biological constraints helps explain similarities and differences in real-life food choices. All of the data needed to compute and visualize substitution and complementarity between these three foods is provided in Table 1.

**Table 1. Cost and composition (vitamin A, iron, and energy) of black beans, butternut squash, and corn flour**

| | Food cost (per package and per serving) | | | | | Food composition (per serving) | | | |
|---|---|---|---|---|---|---|---|---|---|
| | Price per package (USD) | Quantity per package | Serving size (U.S.) | Serving size (grams) | Price per serving (USD) | Vit. A (mcg RAE) | Iron (mg) | Energy (kcal) | USDA FDC ID |
| Black beans, canned | 2.49 | 29 oz can | 0.5 cup | 143 | 0.36 | 0 | 2.71 | 130 | 175188 |
| Butternut squash, diced | 4.99 | 48 oz pkg | 1 cup | 140 | 0.51 | 745 | 0.98 | 63 | 169295 |
| Corn masa flour | 4.99 | 4.4 lb | 1 cup | 122 | 0.33 | 0 | 2.90 | 440 | 169748 |

Notes: Data on prices, package and serving sizes were downloaded from stopandshop.com. Data on the nutrient composition per serving of each item were downloaded from the USDA FoodData Central (FDC) repository (https://fdc.nal.usda.gov), matching items to the USDA FDC ID number shown. Nutrient requirements as specified in U.S. Dietary Reference Intake levels are lower and upper bounds of 18 and 45 mg/day of iron, and 700 and 3000 mcg/day of vitamin A in retinol activity-equivalent (RAE) units, and for an average healthy 30 year-old female total energy needs might be 2,330 kcals/day.

The data in Table 1 are sufficient to provide a biologically and economically accurate example of linear programming to identify least-cost diets for nutrient adequacy, and compare the result to what people actually consume. Graphical representation of this model is shown in Figure 1, which can be used by instructors as a reference from which to build whiteboard sketches or their own slides. For visualization it is helpful to focus on just the two nutrient-rich foods, black beans and butternut squash, and begin with their total cost per day using prices from Table 1:

*(1)*     $C = 0.36x + 0.51y$

In this equation, $C$ is the diet's total cost in US dollars per day, $x$ and $y$ are the quantities of black beans and butternut squash respectively in servings per day, 0.36 is the price per serving of black beans, and 0.51 is the price per serving of butternut squash. Students can be coached to draw cost lines by solving for $y$, so they see that equation (1) can also be written as $y = C - (0.36/0.51)x$. The intercept of each cost line on the $y$ axis is total diet cost in terms of that food (butternut squash), and each cost line has as its slope the real price of the $x$ axis food (black beans) in terms of $y$. Instructors can use this line to discuss opportunity cost, marginal rates of substitution, indifference curves and how revealed preferences explain the quantities that consumers actually choose, but instructors can also proceed directly to identify how biological constraints would affect the quantities needed to meet nutrient requirements for health.

As shown in Table 1, to achieve nutrient adequacy for a representative 30 year-old adult woman, total iron must be within 18 and 45 mg/day, and vitamin A must be within 700 and 3000 mcg/day in retinol activity-equivalent (RAE) units, as computed from Dietary Reference Intakes produced for the U.S. and Canada by the National Academies' Institute of Medicine (Otten, Hellwig, and Meyers 2006). Each constraint can readily be drawn in terms of $x$ and $y$, adding that quantities cannot be negative numbers, from the following equations:

*(2)*     $2.71x + 0.98y \geq 18$ (*total iron must be $\geq$ 18mg/day, from beans plus squash*)
*(3)*     $2.71x + 0.98y \leq 45$ (*total iron must be $\leq$ 45mg/day, from beans and squash*)
*(4)*     $745y \geq 700$ (*total vitamin A must be $\geq$ 700mcg/day, available only from squash*)
*(5)*     $745y \leq 3000$ (*total vitamin A must be $\leq$ 3000mcg/day, available only from squash*)
*(6)*     $x, y \geq 0$ (*the chosen quantities of each food must be non-negative*)

Students can be coached to draw each constraint by solving for y in terms of x, or instructors can go directly to a visual representation of these constraints as shown in Panel A of Figure 1.

**Figure 1. Visualizing least-cost nutrient adequate diets in two dimensions**

Panel A. Adequacy of two nutrients (vitamin A and iron) from two foods (beans and squash)

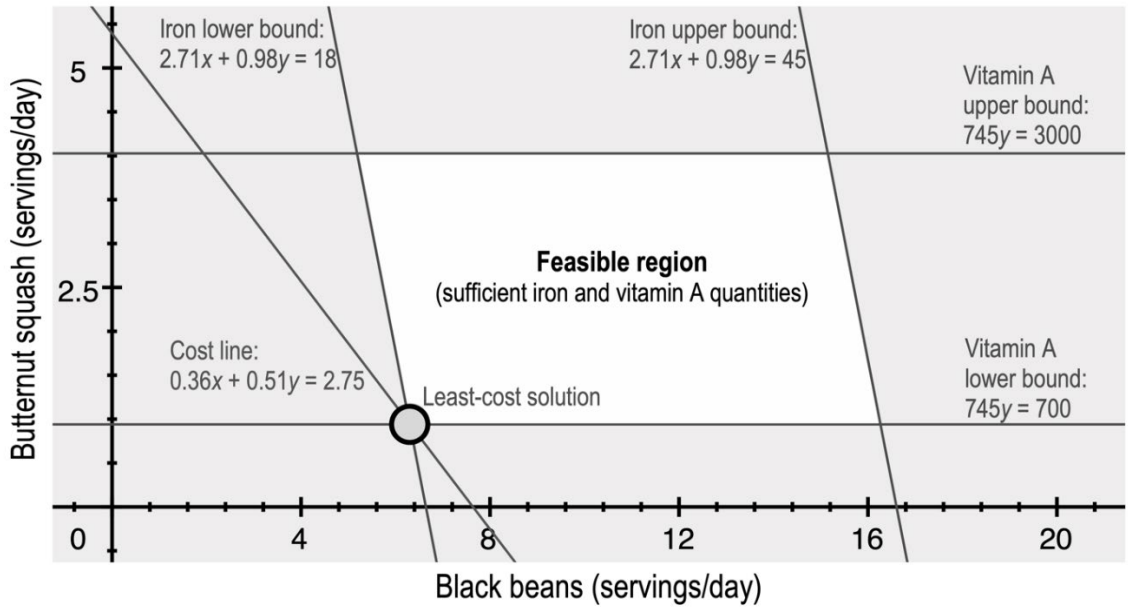

Panel B. Adequacy of vitamin A, iron and energy from beans and squash, with sufficient corn

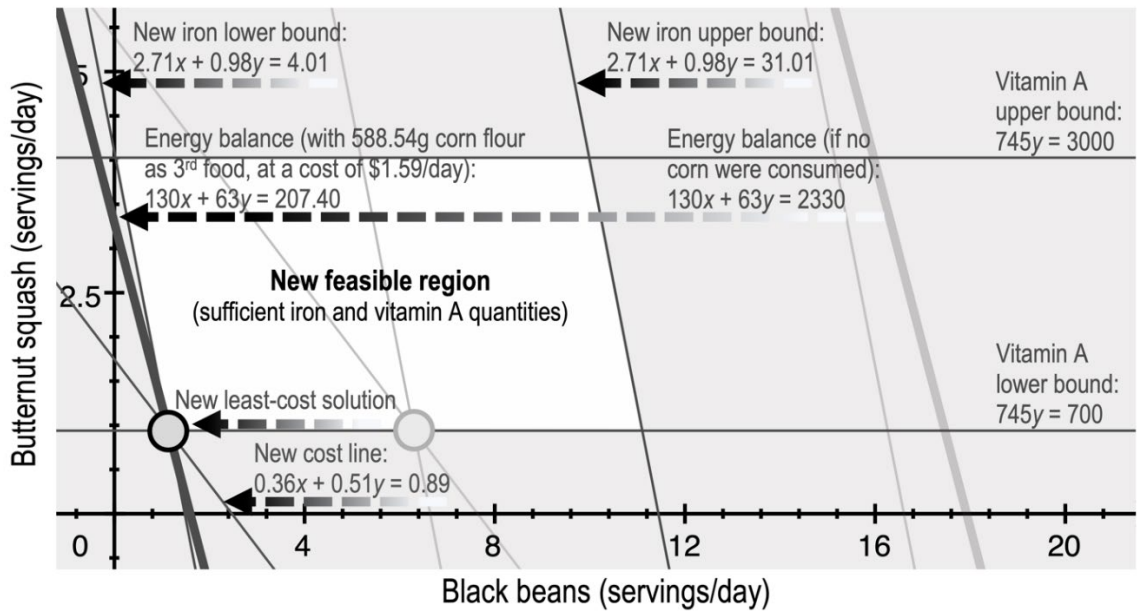

Notes: Panel A illustrates all possible combinations of black beans and squash that satisfy iron and vitamin A constraints, and identifies the least-cost solution among these options. Panel B shows how adding in the least-cost level of corn flour causes the iron and energy constraints, and also the least-cost solution, to shift. Black bean, butternut squash, and corn flour prices per serving were calculated using data from stopandshop.com. Data on the iron, vitamin A, and energy content per serving of black beans, butternut squash, and corn flour were downloaded from the USDA FoodData Central at https://fdc.nal.usda.gov. Iron and vitamin A constraints were obtained from the Dietary Reference Intakes for the U.S. and Canada, and the energy requirement is for 30 year-old females, based on the World Health Organization (WHO) reference end-growth median heights and weights (weight=57 kg, height=163 cm, and maternity status=not pregnant or lactating) and on an 'active' level of physical activity.

Instructors can readily show how the parameters of each constraint in Figure 1 come from the table of data, and can use analysis of units to confirm that quantities are accurately represented. The coefficient 2.71 is the iron content of black beans (in mg/serving), the coefficient 0.98 is the iron content of butternut squash (in mg/serving), and 745 is the vitamin A content of butternut squash (in mcg/serving), while vitamin A content of beans is negligible and dropped from those equations. Even without algebra, students can readily see how nutrient constraints form a biologically plausible parallelogram, imposing upper and lower bounds on the quantities of squash and beans needed to avoid diseases such as iron-deficiency anemia, loss of eyesight due to vitamin A deficiency, or excess levels of these nutrients.

Once the parallelogram in Panel A is drawn, students can imagine that for low-income populations the relevant constraints are lower bounds, for which the least-cost solution would be the intersection of equations (2) and (4). Approximate quantities are clear from visual inspection, and exact values can readily be computed by solving the system of two equations to show that the least-cost solution is $x = 6.30$ servings per day of black beans and $y = 0.94$ servings per day of butternut squash. Equation (1) reveals that this diet would cost $2.75 per day, but attentive students might notice that this amount of food is much less than they would need to survive. This diet meets two important nutrient needs, but provides less than half of the total energy requirements for a typical adult.  As revealed by this step, the need to maintain energy balance plays a crucial role in nutrient adequate diets.

### *Adding the energy constraint needed for calorie balance*
From Table 1, total energy needed for a representative 30 year-old female is 2,330 kcal/day, which provides a seventh kind of constraint:

*(7)*        $130x + 63y = 2,330$ (*total energy must equal 2,330kcal*)

Here the coefficients 130 and 63 are the energy content of black beans and butternut squash respectively, in kcal/serving. Unlike the upper or lower bounds on individual nutrients, energy balance imposes an equality constraint to avoid unwanted weight gain or loss.  The target shown is for a physically active 30 year-old female, not pregnant or lactating, at the median height and weight of a healthy reference population as compiled by the World Health Organization (WHO Multicentre Growth Reference Study Group. 2006). Other levels of height, weight, and physical activity would lead to different targets for energy balance for people in each age and sex group, based on metabolic relationships that any student can readily obtain for themselves from online calculators (Hall et al. 2011, NAL 2023, NIH 2012).

The energy constraint is introduced at the far right of Panel B of Figure 1, obtained by solving equation 7 for $y$ to obtain its slope of 2.06 (=130/63) servings of butternut squash to obtain the energy from one serving of black beans, and also solving for $x$ to obtain its horizontal intercept at 17.92 (=2,330/130) servings of black beans to attain energy balance when only black beans are consumed. This energy constraint at the far right of Panel B exceeds the upper bound on iron intake for any feasible combination of black beans and butternut squash.

To obtain adequate energy and stay within micronutrient bounds, diets need varying quantities of an energy-dense food in addition to the nutrient-dense items such as beans and squash. In this example we can imagine using corn flour to make tortillas or grits, but other energy-dense foods include any cereal grains and starchy roots, sugar, and vegetable oil or animal fat. If the energy-dense food provided "empty" calories with no iron or vitamin A, energy balance would be met with 3.30 servings of corn flour for the 1,451.78 kcal remaining of the person's total energy needs (2,330 kcal) after subtracting the energy in the least-cost quantities of beans and squash (878.22 kcal/day, from 6.30×130 kcal/serving of black beans, and 0.94×63 kcal/serving of butternut squash).

As shown in the data in Table 1, white corn has no vitamin A but brings significant amounts of iron (2.90 mg/serving) and therefore adds to the iron in beans. Including this third food introduces another dimension to Figure 1, which could be imagined as a third axis labeled z with varying quantities of corn that we project onto $x$ and $y$ in Panel B as an additional thicker line that is colored red. The energy constraint with zero corn was drawn at the far right of Panel B. Since corn brings both energy and iron, raising the level of corn above zero shifts both the energy constraint and the feasible region to the left. Each additional serving of corn shifts the energy constraint leftward by 3.38 servings of black beans (440/130), and shifts the iron constraint leftward by 1.07 servings of beans (2.90/2.71), because corn flour is relatively more energy-dense than black beans. One could imagine drawing Panel B with sufficient corn flour to meet all of the iron requirement, but that would require 6.21 servings (to deliver 18 mg/day of iron at 2.90 mg per serving of corn flour), and when combined with enough butternut squash for vitamin A there would be excess energy at 2,790 kcal/day. Staying within the energy boundary at 2,330 kcal/day requires a mix of corn, beans and squash. The actual energy requirement from corn and beans depends on the quantity of squash, which is the only source of vitamin A, so between 0.94 and 4.03 servings are needed to stay between the 700 to 3,000 mcg RAE limits at 745 mcg RAE/serving. Each level of squash would also bring between 59 and 254 kcal/day, given its energy content of 63 kcal/serving, leaving between 2,076 and 2,271 kcal/day to be provided by corn and beans.

The exact combination of corn, beans and squash that would meet all constraints depends in part on their relative prices shown in Table 1. Among these three foods, butternut squash is by far the most expensive source of energy, at $0.81 per hundred kcal, from $0.51/serving divided by 63 kcal/serving, in contrast to beans at $0.28 per hundred kcal from $0.36/serving divided by 130, and corn at $0.08 per hundred kcal from $0.33/serving divided by 440. These extreme cost differences ensure that vitamin A from butternut squash is a binding constraint, at the lower bound of butternut squash which leaves 2,270.8 kcal/day to be provided by corn and beans. That is the lower-left corner of the trapezoidal feasible region in Panel B, which is the solution to two equations in two unknowns, which can be solved to reveal the need for 4.82 servings of corn flour and 1.14 servings of black beans, in addition to the 0.94 servings of butternut squash.

***Generalizing to more foods, more nutrients, and observed variation in the real world***
The two-dimensional diet problem shown in Figure 1 provides an accessible demonstration of how linear programming for nutrient adequacy generates model-based benchmarks of relevance in the real world. In this and other modeled diets, foods complement each other such that nutrient-adequate diets use them in fixed proportions. The same constraints often remain binding as prices change, so substitution mainly occurs between foods with very similar nutrient density. The large quantity required of the starchy staple ensures that it accounts for most of the total cost (in this case $1.59 per day, $0.33×4.82), but the high cost of producing and distributing nutrient-rich complements makes them expensive (in this case $0.89 per day, from $0.36×1.14 servings of black beans plus $0.51×0.94 servings of butternut squash), for a total cost at these prices of $2.48 per day.

The simple least-cost diet with only three foods shown in Figure 1 uses a small subset of the data in the full exercise, and represents a tiny fraction of all modeled diets ever computed for research and policy documents, but provides a surprisingly powerful explanation of real-life food choice and its consequences. The full exercise described in this paper offers 60 different foods at prices observed in Boston in November 2023, with which to meet upper and lower bounds of 22 nutrients for which Dietary Reference Intakes have been established (NASEM 2019). That dimensionality reveals important roles for low-cost plant oils, dairy, and a variety of vegetables, but only 8-11 items are needed to meet all constraints, and the additional expense of those additional items is modest at a total cost of $2.88 per day.

The stylized models in this exercise use a single set of food prices in Boston and a single representative person, but similar results are found in ongoing research that accounts for differences in food composition, market prices, and nutrient requirements. A key finding is the narrow range of variation in least-cost nutrient adequate diets, as nutritionally similar items substitute for each other with relatively small differences in the total cost of production and distribution of enough locally available foods to meet nutrient needs. Using actual food price data from 2017 across all countries of the world at each age and sex, the global median least-cost nutrient adequate diet adds up to $2.32 per person at PPP prices (Bai, Herforth, and Masters 2022), with an interquartile range from $1.95 to $2.76.

Least-cost nutrient adequate diets computed for populations worldwide are dominated by locally available energy-dense staples like corn, with small quantities of the more expensive locally available nutrient-dense items like squash or beans, in quantities and costs per day that are similar to those calculated in this classroom exercise. These diets are nutrient adequate but lack sufficient quantities of desirable food groups such as vegetables and fruits to meet the Dietary Guidelines for Americans (HHS and USDA, 2020) or other nutritional recommendations. To measure whether vegetables, fruits and other healthy foods recommended in dietary guidelines are affordable, Masters et al. (2018) introduced price indexes based on least-cost items by food group. Adding up least-cost items in proportions recommended by dietary guidelines leads to a daily cost of healthy diets used to monitor global food security by international agencies (FAO, IFAD, UNICEF, WFP and WHO 2020, 2021, 2022, 2023). As shown in the resulting Food

Prices for Nutrition database maintained by the World Bank (2023), and also in FAOSTAT (FAO 2023), using the most recent internationally comparable PPP prices for 2017, the global average cost per day of items for a healthy diet was $3.30.

Comparing least-cost diets that provide increasing levels of more expensive attributes, the World Bank's Food Prices for Nutrition database shows the cost per day of just enough dietary energy from each country's least expensive starchy staple to be $0.83, in contrast to the comparable global average cost of nutrient adequacy of $2.44, or as mentioned above the cost of a healthy diet at $3.30. This ladder of least-cost diets, from a subsistence floor for dietary energy that costs under $1/day, to cost of nutrient adequacy in the $2-3 range, and a cost of healthy diets in the $3-4 range, shows how the least-cost diet exercise that introductory economics students can do in a classroom is very closely related to fundament work on economic policy, agricultural development and global health.

When teaching this module, moving from the three-food example to the full exercise in this paper adds dimensionality without altering basic principles illustrated in Figure 1 and demonstrated computationally in Excel. Most importantly, students can immediately see that many people (including themselves) might consume less of the low-cost nutrient-rich foods like squash and beans than would be needed for nutrient adequacy, either because their incomes are so low that they need to meet energy needs with only corn and other even lower-cost foods, or because meal preparation constraints and consumption preferences lead to meeting energy needs with other items.

### *Comparing least-cost diets to actual consumption*
A key feature of the exercise is comparing least-cost diets to actual consumption. Students can appreciate that nutrient adequacy is not required for short-term survival, but many food traditions reflect the health benefits of nutrient-rich foods. Indeed, the three items in Figure 1 were chosen to represent the "three sisters" of Mesoamerica (Marsh 2021, Ngapo et al. 2021). Combining corn, beans, and squash or similar crops takes advantage of agronomic complementarities when grown together, and offers nutritional complementarities in consumption as shown in our modeled diets. Traditional diets included the three sisters in quantities similar to the least-cost nutrient adequate levels found in Panel B in Figure 1, as revealed for example by a 1971 survey of mothers in the village of Santa María Cauqué, Guatemala (Mata 1978, page 106). In that study, adult women at the start of pregnancy were found to eat primarily corn tortillas (621g/day) with beans (53g/day) and vegetables (46g/day), which is somewhat less than the 589, 163 and 132 g/day needed to reach the levels of energy, iron and vitamin A shown in Figure 1. These very low-income Maya women in 1971 had diets that were similar but different from our least-cost nutrient adequate benchmark in important ways: for example, in addition to the basic staples they consumed about 15 teaspoons (61g) of sugar per day, and their primary dietary deficiency was insufficient intake of vitamin A-rich vegetables like butternut squash. Students with a very wide range of food preferences and cultural backgrounds can appreciate the universal relevance of least-cost nutrient adequate diets as a benchmark that meets constraints for lifelong health, which actual diets often fail to achieve for a variety of reasons.

**Excel workbook for calculating least-cost diets with local food items and prices**
The Excel workbook and exercise instructions provide self-contained materials with which to compute the combination of foods that meet, or come close to meeting, a specified set of nutrient constraints at the lowest possible cost, and compare that least-cost diet to actual food choices.

The food items from which the least-cost diet is computed are a selection of 60 food items available for delivery from the Stop & Shop online grocery store (via peapod.com) to the Tufts University campus in Boston, Massachusetts in November 2023. The food list includes a variety of commonly consumed food items that one might expect to see in everyday use. Foods were grouped into five categories for ease of comparison among products that might substitute for each other, as: 1) starchy staples like corn; 2) fruits & vegetables like squash; 3) nuts, beans, seeds, & oils; 4) animal-source foods and alternatives; and 5) milk & nutrient-dense beverages. Each item was matched to its description in the USDA National Nutrient Database for Standard Reference 28 (USDA 2018), the USDA Global Branded Food Products Database (USDA 2023a), or the USDA Foundation Foods Database (USDA 2023b). Users of the exercise can readily update the food options, entering the nutrient composition of any additional items obtained from the USDA's online portal, to reflect the items and prices available at any place and time.

Requirements use Dietary Reference Intake (DRI) data[1] from the Institute of Medicine of the National Academies of Science, Engineering and Medicine (Otten, Hellwig, and Meyers 2006; NASEM 2019) compiled for energy and a selection of 22 nutrients, following recommendations outlined in Schneider and Herforth (2020), with some modifications made to best suit this activity[2]. Recommended dietary allowance (RDA) or adequate intake[3] (AI) are included in the model as lower bounds for micronutrient intake levels. Tolerable upper intake levels (UL) are applied as upper bounds, where applicable. The chronic disease risk reduction (CDRR) level established for sodium in 2019 is included as the upper bound for sodium intake. The acceptable macronutrient distribution ranges (AMDR) for protein, total fat, and carbohydrates are included as their respective lower and upper bounds. The AI for fiber of 14g per 1000kcal was converted to grams per day and is included as a lower bound[4]. The estimated energy requirement (EER) was obtained from the USDA DRI calculator for healthcare professionals (at https://www.nal.usda.gov/human-nutrition-and-food-safety/dri-calculator) and set to maintain daily energy balance with no weight gain or loss, and thus represents both a lower and upper bound. These data were compiled for the population of 30 year-old adults, based on the World Health Organization (WHO) reference end-growth (19 years of age) median heights and weights (for a woman at weight=57 kg, height=163 cm, and maternity status=not pregnant or lactating, and a man of weight=67 kg, height=177 cm) and on an 'active' level of physical activity, which is the level recommended for long-term health (de Onis et al. 2007; WHO Multicentre Growth Reference Study Group 2006). Energy requirements were rounded to three significant digits, at 2,330 kcal/day for women and 2,900 kcal/day for men.

All data needed for the exercise are contained in one Excel workbook, in tables labeled '1.FoodPricesAndComposition' and '2.NutrientRequirements'. The Excel workbook also contains worksheets formatted to allow students to make use of these data to build diets for females and males in tables labeled '3.Guesswork1-female', '4.Guesswork2-female', '5.Guesswork1-male', '6.Guesswork2-male'. Students are asked to experiment with different diet plans to try to guess at the least-cost diet plan that meets targets for energy and all 22 nutrients. The female and male guesswork spreadsheets permit users to input values to designate any number of servings and any combination of food items (see column in yellow, headed 'Servings'). The Excel workbook is designed so that total energy and nutrient adequacy is updated immediately and shown to the student in a bar graph and also numerically.

### Guesswork with visual feedback on proximity to model constraints

In this exercise, when students choose items and enter its number of servings, Excel provides immediate visual feedback on the degree to which nutrient requirements are met. Bar graphs show the diet's energy content and nutrient levels relative to their upper and lower bounds for health, and color-coding of the cells showing each numerical result indicate whether the diet delivers each nutrient within or at each threshold. Cells display the percent attained of the lower (row 8) and upper bounds (row 9), turning a darker shade of red when values are further outside either boundary, a darker shade of blue when values are closer to the midpoint between upper and lower bounds, and nutrient levels at exactly 100 percent of a boundary are white with bold text. A diet meeting all nutrient targets would not show any red-colored cells, but would show cells in various shades of blue and some white, bolded cells.

### Comparing guesswork diets to the solved diet for nutrient adequacy at least cost

After students experiment with guesswork, instructors can reveal how Excel computes an exact solution with results shown in worksheets labeled SOLVED-female and SOLVED-male. It turns out that a small set of foods (11 for the representative female, 8 for the representative male) are sufficient to meet all constraints. By the fundamental theorem of linear programming, that same number of constraints (either 8 or 11) are binding in each problem, while all other nutrients will be within bounds and therefore shaded blue. Students can see how the model was specified in Excel's Solver add-in, and instructors who want to explain the mathematics can spell out the problem on a whiteboard as:

$$Min\ C = \sum_i p_i * q_i$$

Where,

$$q_1 \geq 0, q_2 \geq 0, q_3 \geq 0, \dots, q_i \geq 0\ , \qquad i = 1, \dots, 60$$

and

$$\sum_i a_{ie} * q_i = EER$$

and

$$\sum_i a_{ij} * q_i \geq LB_j\ , \qquad j = 1, \dots, 21$$

and

$$\sum_i a_{ik} * q_i \leq UB_l \,, \qquad l = 1, \dots, 16$$

The model's objective is to minimize diet cost $C$, obtained by multiplying each item's price, $p_i$ by its quantity consumed, $q_i$, and summing across all food items, given four kinds of constraints: (a) quantities chosen cannot be less than zero, (b) the sum of energy provided by each food, $a_{ie}*q_i$, must be equal to the person's estimated energy requirements (EER) to avoid unwanted weight loss or gain; (c) the sum of nutrients for which lower bounds have been established, $a_{ij}*q_i$, must equal or exceed those lower bounds to avoid deficiency diseases, and (c) the sum of nutrients for which upper bounds have been established, $a_{ik}*q_i$, must be equal to or less than those upper bounds to avoid toxicity. The lower bounds (*LB*) are nutritionally defined as either an RDA, an AI, or the lower bound of the AMDR depending on the nutrient in question, and upper bounds (*UB*) are nutritionally defined as either a UL, a CDRR, or the upper bound of the AMDR.

Figure 2 shows how to install the Solver add-in and specify the model for it to compute each least-cost diet using the simplex algorithm. Figure 3 shows how bar charts embedded in the Excel workbook will appear when the female and male diet problems have been solved.

**Figure 2. Solving the least-cost diet problem with Excel's Solver**

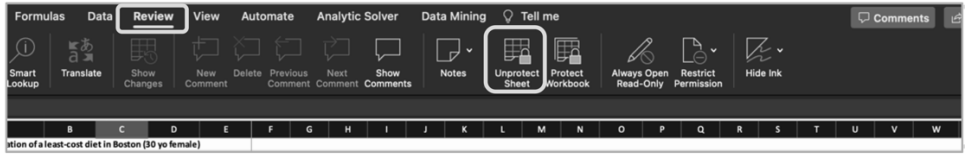

- Under the Review tab, select 'Unprotect Sheet'.

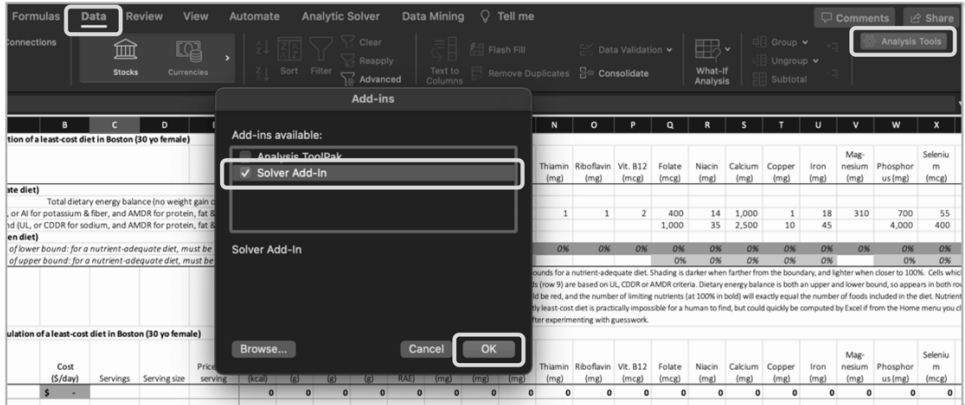

- Under the Data tab, select 'Analysis Tools'.
- In the 'Add-ins' pop-up window, select 'Solver Add-In', then click 'OK'.

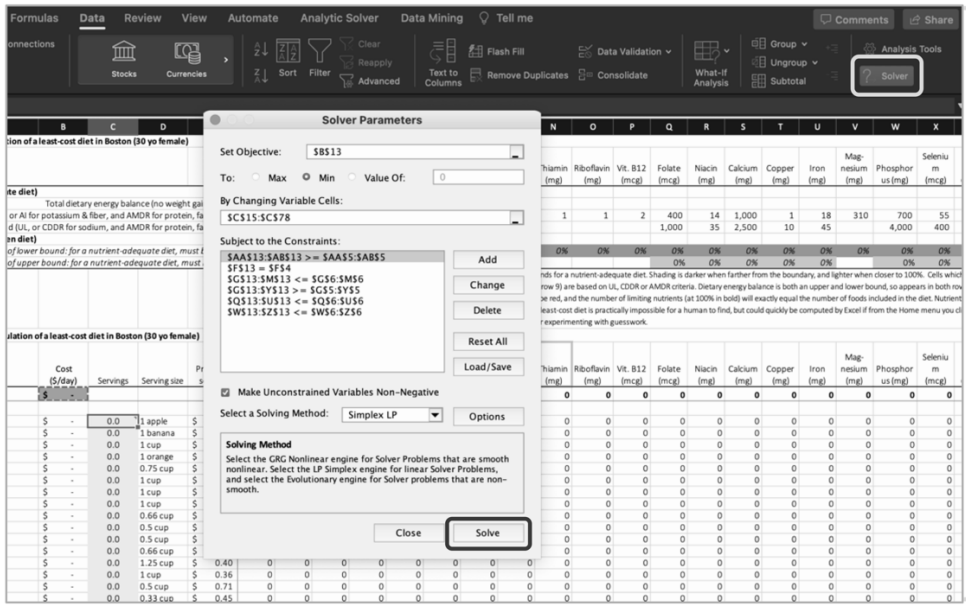

- In the top right corner, click 'Solver'.
- Excel will have remembered the necessary parameters shown in the 'Solver Parameters' pop-up window. Note that the objective is to minimize the total cost in cell B13 by varying the number of servings in cells C15-C78, while satisfying energy (row 4) and nutrient constraints (rows 5 & 6).
- Click 'Solve' in the bottom right corner of the 'Solver Parameters' window.

Notes: Images are from the female and male guesswork sheets in the provided Excel workbook, and show how to install and use Excel's Solver add-in.

**Figure 3. Visual representation of energy and nutrient levels in solved least-cost diets**

Panel A. Percent attained of nutrient lower and upper bounds in solved least-cost diet (30 yo female)

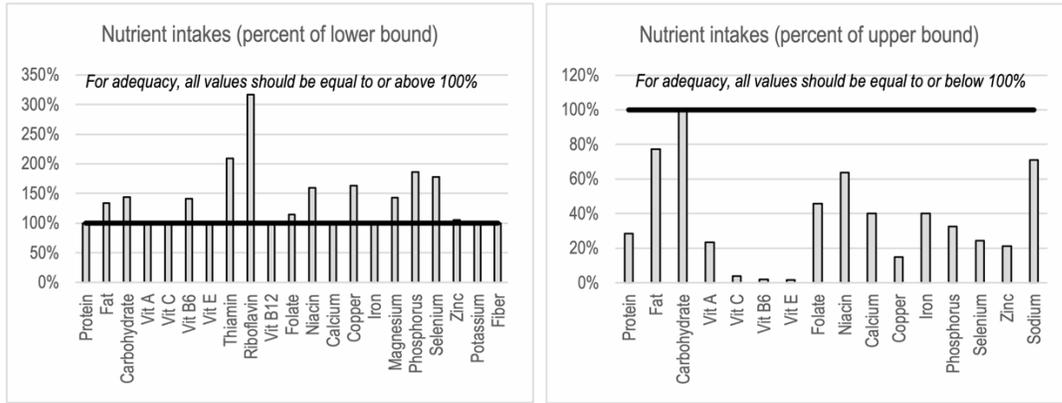

Panel B. Percent attained of nutrient lower and upper bounds in solved least-cost diet (30 yo male)

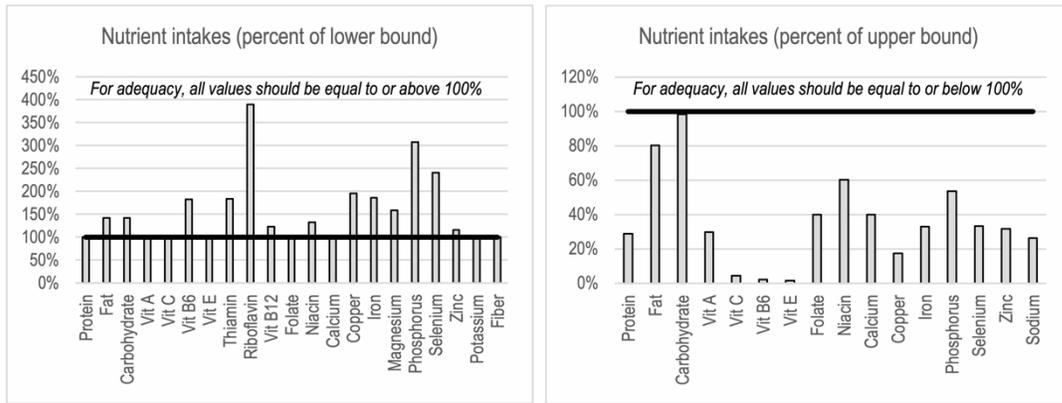

Panel C. Percent attained of energy requirement in solved least-cost diet

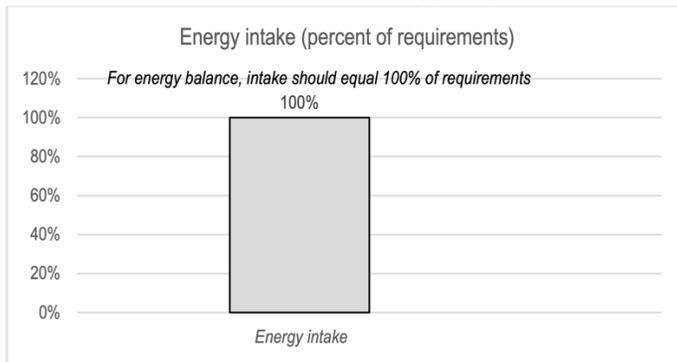

Note: Bar charts are from the "10.SOLVED-female" and "11.SOLVED-male" tabs of the Excel workbook.

***Exploring change in solved diets due to changes in price or other parameters***
If instructors or students wish to explore the effect of price or other changes on least-cost diets, they can easily change the price of a food item in the sheet labeled "2.FoodPricesAndComposition", which will update relevant cells in column E of the guesswork and solved sheets. For example, if the price of fat-free milk were to double from $0.30/serving to $0.60/serving, the solved least-cost diet for females shifts from $2.88/day to $2.89/day, fat-free milk is no longer part of the diet, and whole milk is included. Likewise, the solved least-cost diet for males increases from $3.17/day to $3.21/day, and fat-free milk is replaced by whole milk. Instructors and students can readily experiment with all kinds of changes in model parameters, exploring when those lead to no change at all in the foods used for a least-cost diet, or substitution among similar items that meet the same nutrient constraints, or sufficient change in the cost of nutrients such that different constraints become binding.

***Summarizing output from the least-cost diet Excel workbook***
Following their guesswork, students are tasked with preparing a report that includes tables to display the results of two distinct male and female diet plans, and the solved least-cost diets. In the first table, students are asked to show the quantity of each food in the four guesswork diet plans and their total cost per day, as well as the corresponding result obtained using mathematical programming by Excel's Solver. In a second set of tables, students are asked to show the quantity of each nutrient in each diet, and their adequacy as a percentage of the lower or upper bound. In a third table, students can show the nutrient composition of the solved least-cost diets. These tables have already been constructed for students and can be found in the orange tabs of the Excel workbook labeled "7.Table1-ForReport", "8.Table2-ForReport", and "9.Table3". Their contents will fill automatically from the data in the other tabs, including the quantities of each food chosen in the Guesswork sheets. Students are also asked to address several additional questions intended to facilitate further exploration of concepts related to nutrient adequacy and diet cost, which are provided in Appendix A.

**Classroom takeaways**
This teaching module was developed for the start of an introductory course on the economics of food and nutrition taken primarily by students earning M.S. degrees related to food policy in the School of Nutrition at Tufts University, and would be applicable to any graduate or upper-level undergraduate economics course. About a third of students in this course have never taken any economics course at all, and many report having actively avoided taking economics due to concerns about mathematical abstraction and focus on business and finance. Doing the exercise fosters discussion during and after class on variety of topics as described below. Further context is provided in the supplemental materials online, as Appendix A for instructors and Appendix B for students, in addition to Appendix C which is the Excel workbook itself.

A common first reaction among students is the surprising nature of the task: comments in our online discussion forum from Fall 2022 included "this was a mind-blowing exercise", "a new experience", "an eye-opening, hands-on opportunity to see what it truly costs to purchase the bare minimum", and "an insightful experience on many levels". Students valued the playful quality of guesswork, one describing it as a "fun and interactive way to merge economics and nutrition" that "felt like a game", and another as "both enjoyable and informative - unlike anything else I've ever done in a classroom setting". Some focused on surprising nature of the computational challenge, noting that "it was shockingly difficult to construct a diet that both met the nutrient requirements and remained low-cost", and they appreciated doing the exercise together and discussing results in class, noting that "The activity itself - even for as brief as a time as we did it - was so amazingly helpful and really helped me grasp just HOW difficult and WHY it was so difficult."

A second frequent theme of student comments is how the exercise helped them imagine someone else's food choices, describing it as "a tangible exercise that will help tremendously with empathy". One student said "I enjoyed doing it and was super interested to see the solution of pasta, oil, broccoli and canned spinach", which happened to be among the least cost items in their year. Some wrote about how their everyday experience with foods advertised as healthy led them to think that sufficient nutrients would be much more expensive than the solved diets. For example, writing about the role of vegetable oil, a student noted that the exercise made them realize how often nutrition advice "does not account for income and restricted budgets, and it feels irresponsible for public figures to be demonizing a food (like vegetable oil) that is a necessity for many individuals.' Others noted with interest how "consuming a nutrient adequate diet does not have to be expensive" and "can be achieved by less than $4 a day" in the U.S., even as low-income people in countries like Ethiopia often cannot afford nutrient adequate quantities of the foods available in their local markets.

Third, students quickly recognized that achieving nutrient adequacy is not just a matter of having enough money. Students discovered how difficult it is to know which grocery items would make up a nutrient adequate diet, and also noticed the major role of other factors that drive food choice such as time use in meal preparation, preferences and aspirations beyond nutrient adequacy. These observations point to the central role of food product regulation, technology and policy in shaping health outcomes, prompting comments about how "food fortification and enrichment plays a huge role" by helping consumers unknowingly meet more of their nutrient needs with basic staples. Students contrasted the situation in Boston with Ethiopia, where those same foods are often not fortified, refrigeration is rarely available, and cooking technology as well as time use and income constraints contribute to food choices such as use of dry beans and lentils.

Finally, at nutrition school many students know a lot about the biology of food and health, and find it difficult to use a highly stylized model that focuses only on nutrient adequacy. As one student wrote "it is challenging for me to ignore dietary guidelines and overall long term dietary considerations". Common suggestions for additional health constraints include targets for entire food groups or phytochemicals and other health-promoting components in fruits and vegetables, as well as target ratios of healthy to unhealthy forms of each macronutrient, especially healthy fats and whole grains. Students also call for costing the time needed for meal preparation or imposing constraints on time and cooking technology, as well as nutritional constraints for special needs such as gluten-free or low-sodium diets. Finally, many suggest constraints that reflect a person's values such as vegan or vegetarian diets, and religious or cultural norms such as halal and kosher certification or regional and ethnic dietary patterns. All of these reveal how a simple least-cost diet model can be used as a starting point to frame discussion, identifying which aspects of consumer behavior omitted from the initial model could be represented in the objective function or constraints, and the extent to which revealed preferences reflect the nutrients required for a person's present and future health.

**Conclusion**

This paper describes an Excel workbook and associated exercise through which students without mathematical preparation can observe the cost and nutrient adequacy of alternative food choices, experiment with guesswork about daily diets that would meet nutrient requirements at affordable prices, and then see the exact least-cost diet for nutrient adequacy using Excel's Solver. The accompanying guide for instructors and students can be used online or in-person, as a standalone activity or linked to other teaching about consumer behavior and utility maximization in economics, or about optimization and linear programming in other fields. The basic activity can be done quickly in introductory courses, and the same Excel workbook can be explored in greater depth for further work on how the affordability of essential nutrients relates to health disparities, poverty measurement, economic history and international development.

For instructors who want to introduce the exercise graphically, we provide actual data for drawing a cost line subject to piecewise linear nutrient constraints, and show how that relates to a budget line and convex indifference curves used to explain consumer behavior. All parameters of the graphical example and the larger Excel workbook can be readily modified, providing an attractive and accessible way to explore how changes in food consumption alter cost and nutrient adequacy, and how the least-cost solution varies with price, food composition and nutritional needs.

Ever since the early 20th century discovery of essential nutrients, least-cost diets have been used for instructional purposes and to inform nutritional interventions. More recently, least-cost diets have been used as a new kind of price index, to calculate whether people can afford enough of the locally available foods to meet their nutrient requirements for present and future health. The Excel workbook and instructional notes provide an accessible, and empirically accurate way for students to use updated data in visual form, with guidelines and a Word template for assignment submission to elicit insights about how least-cost diets relates to consumer behavior, health equity and economic development. These materials can help introduce students to the use of optimization models and also be used in more advanced courses, research projects and policy analyses.

**Endnotes**

[1] Dietary Reference Intakes consist of four types of reference values: (1) the estimated average requirement (EAR); (2) the recommended dietary allowance (RDA); (3) the adequate intake (AI); and (4) the tolerable upper intake level (UL). The Institute of Medicine of the National Academies has also established acceptable macronutrient distribution ranges (AMDR) for each of the macronutrients, as well as a chronic disease risk reduction (CDRR) level for sodium.

[2] Schneider and Herforth (2020) recommend including only micronutrients with an established EAR (which will also have a set RDA). For this activity, we use RDAs where available, but we also include the AI for potassium, for which there is no established EAR or RDA. Additionally, we include the AI for the macronutrient, fiber.

[3] The RDA defines the intake value considered necessary to meet the needs of 97-98% of a healthy population. Where there is insufficient evidence to set an RDA, an AI may be established. An AI provides an approximation of the intake level expected to meet or exceed the nutrient needs of most healthy individuals in a demographic group.

[4] The AI for fiber of 14 g per 1000kcal was converted to grams per day using our calculated energy requirements for healthy populations of 30 year-old females and males (females: 14g ÷ 1000kcal * 2330kcal = 33g fiber per day; males: 14g ÷ 1000kcal * 2900kcal = 41g fiber per day).

**Data availability statement**

The graphical example and Excel workbook in this article were built with data from the public domain, using sources detailed in the text. As with other least-cost diets research, our graphical model and the larger Excel version combine data on food prices, which in this case are drawn from the Stop & Shop grocery store via peapod.com, linked to the nutrient composition of those items from the USDA FoodData Central repository at https://fdc.nal.usda.gov, and estimated requirements for individual nutrients from the National Academies' Dietary Reference Intakes (DRI) values at https://nap.nationalacademies.org/collection/57/dietary-reference-intakes, as well as overall energy balance for a representative 30 year-old person based on the World Health Organization reference end-growth median heights and weights and an active level of physical activity (de Onis et al. 2007; WHO Multicentre Growth Reference Study Group 2006), calculated using the USDA DRI calculator for healthcare professionals at https://www.nal.usda.gov/human-nutrition-and-food-safety/dri-calculator.